\documentclass[aps,prc,twocolumn,showpacs,preprintnumbers,amsmath,amssymb]{revtex4}
\usepackage{graphicx}
\usepackage{graphicx}
\usepackage{dcolumn}
\usepackage{bm}

\newcommand{\be}{\begin{eqnarray}}
\newcommand{\ee}{\end{eqnarray}}

\newcommand{\vav}{\left\langle v_2\right\rangle}

\begin{document}


\title{Does the NJL chiral phase transition affect the elliptic flow of a fluid at fixed $\eta/s$? }

\author{S. Plumari $^{1,2}$}
\author{V. Baran $^{2,4}$}
\author{M. Di Toro$^{1,3}$}
\author{G. Ferini}
\author{V. Greco $^{1,3}$\email{greco@lns.infn.it}}
\affiliation{$^1$ Dipartimento di Fisica e Astronomia, Universit\'a di Catania, Via S.Sofia 64, 95125 Catania, Italy\\
$^2$INFN, Sezione di Catania, 95125 Catania, Italy\\
$^3$INFN-LNS, Laboratori Nazionali del Sud, Via S.Sofia 62, 95125 Catania, Italy\\
$^4$Physics Faculty, University of Bucharest and NIPNE-HH, Bucharest, Romania}

\date{\today}

\begin{abstract}
We have derived and solved numerically the Boltzmann-Vlasov transport equations that includes both two-body collisions and the chiral phase transition by mean of NJL-field dynamics.
The scope is to understand if the field dynamics supply new genuine effects on the build-up of the
elliptic flow $v_2$, a measure of the asymmetry in the momentum space, and in particular if it can affect the relation between $v_2$ and the shear viscosity to entropy ratio $\eta/s$.
Solving the transport equation with a constant cross section for the condition of $Au+Au$
collisions at $\sqrt{s_{NN}}=200$ AGeV it is shown a sizable suppression of
$v_2$ due to the attractive nature of the field dynamics that generates the constituent mass.
However the key result is that if $\eta/s$ of the system is kept fixed
by an appropriate local renormalization of the cross section the $v_2$ does not depend on the details of the collisional and/or field dynamics and in particular it is not affected significantly by the chiral phase transition.
\end{abstract}

\pacs{25.75.-q , 25.75.Ld, 12.38.Mh, 24.85.+p }

\maketitle

{\it Introduction - }
The ultra-relativistic heavy-ion collisions at high energy $\sqrt{s_{NN}}\sim 200$ AGeV represent the main tool to study the formation
and the properties of the quark-gluon plasma at high temperature. The RHIC program at BNL has shown that the azimuthal asymmetry in
momentum space, namely the elliptic flow $v_2$, is the largest ever seen in HIC suggesting that an almost perfect fluid with a very small
shear viscosity to entropy density ratio, $\eta/s$, has been created \cite{Adams:2005dq,Adcox:2004mh,Shuryak:2008eq}.
From simple quantum mechanical considerations \cite{Danielewicz:1984ww} as well as from the study of supersymmetric Yang-Mills
theory in the infinite coupling limit a lower bound for $\eta/s$ of about $\sim 10^{-1}$
is predicted \cite{Kovtun:2004de}.
Such a value is much lower than any other known fluid and in particular smaller than the
one of water and even than the superfluid He \cite{Lacey:2006bc}.

First developments of relativistic viscous hydrodynamics \cite{Romatschke:2007mq,Song:2008si} as well as parton cascade models \cite{Xu:2007jv,Molnar:2008jw,Ferini:2008he} indicate that even a small $\eta/s \sim 0.1-0.2$ affects significantly the strength of $v_2(p_T)$ especially at $p_T > 1$ GeV.
Therefore it has become mandatory to determine the value of $\eta/s$ of the plasma created at RHIC
through the study of the relation between $\eta/s$ and $v_2$ \cite{Drescher:2007cd}.
However viscous corrections to ideal hydrodynamics are indeed large and
a simple relativistic extension of first order Navier Stokes equations is affected by causality and stability
pathologies \cite{Romatschke:2009im,Huovinen:2008te}.
It is therefore necessary to go to second order gradient expansion,
and in particular the Israel-Stewart theory has been implemented to simulate the RHIC collisions providing an upper bound for $\eta/s\leq 0.4$
\cite{Song:2008hj}.
Such an approach, apart from the limitation to 2+1D simulations, has the more fundamental problem that it is based on a gradient expansion at second order that is not complete
\cite{Romatschke:2009im}.
Furthermore it cannot be sufficient to describe correctly the dynamics of a fluid with
large $\eta/s$ as the one in the cross-over region and/or hadronic phase which at least at RHIC still gives a non negligible contribution to $v_2$ \cite{Hirano:2005xf} that affects
the determination of the $\eta/s$ itself \cite{Greco:2009ds}.

A relativistic transport approach has the advantage to be a 3+1D approach not based on a gradient expansion that is valid also for large viscosity and for out of equilibrium momentum distribution allowing a reliable description also of the intermediate $p_T$ range where the important properties of quark number scaling (QNS) of $v_2(p_T)$ have been observed \cite{Fries:2008hs}.
In this $p_T$ region viscous hydrodynamics breaks its validity because the relative deviation of the
equilibrium distribution function $\delta f/f_{eq}$ increases with $p_T^2$
becoming large already at $p_T \ge 3T \sim 1 GeV$ \cite{Song:2009rh}.


In this perspective transport approaches at cascade level have already been developed \cite{Xu:2007jv,Molnar:2008jw,Ferini:2008he,Lin:2001zk}, but they miss any effect of the field interactions
responsible for the chiral phase transition or confinement.
With this letter we go one step further including a transport equation self-consistently derived
from the Nambu-Jona
Lasinio (NJL) lagrangian. This allows to study microscopically the transport behavior of a fluid that includes the chiral phase transition
looking at its impact on the relation between the $v_2$ and the $\eta/s$ of the system.
The choice of the NJL is mainly driven by its wide and renowned application to study the QCD chiral phase transition by mean of effective lagrangians, even though the thermodynamical properties of QCD can be reproduced
only qualitatively as briefly discussed in the following.

The NJL Lagrangian is:
\begin{equation}
\label{NJL_Lagrangian}
{\cal L}_{NJL}=\bar{\psi} (i \gamma^{\mu} \partial_{\mu}- \hat{m} ) \psi + g  \bigg[ \Big( \bar{\psi}  \psi \Big)^2 +  \sum_{\alpha=1}^{N_{f}^{2}-1}\Big( \bar{\psi} \tau^{\alpha} i \gamma_5 \psi \Big)^2  \bigg]
\end{equation}
with $\psi$ denoting a quark fields with $N_f$ flavors $\psi=(u,d,...)^t$, $\tau^{\alpha}$ are the generators of the $SU(N_f)$ group acting in flavor space with $\alpha=1,...,(N_f^2-1)$ . The $\hat{m}=diag(m_u,m_d,...)$ is the current $N_f \times N_f$ quark mass matrix in flavor space. In the following we will refer to the $N_f=2, N_c=3$ for calculations. As well know the theory
is non-renormalizable, hence a cut-off $\Lambda$ has to be introduced as a free parameter.
The numerical results shown in the following are derived using the Buballa parametrization: $\Lambda=$ 588 MeV, $g\Lambda^2=2.88$, $m=5.6$ MeV
\cite{Buballa:2003qv} that among the variety of parameterizations entails a behavior
of $\epsilon, P, c^{2}_{s}$ closer to the lQCD results.

A transport theory for the NJL model has been derived in the closed-time-path formalism combined with the effective action method \cite{Zhang:1992rf}. The main steps of derivation are to
perform a Wigner transformation of the Dirac equation of motion and of the related gap-equation associated to
the ${\cal L}_{NJL}$,  Eq.(\ref{NJL_Lagrangian}). Then one exploits the semi-classical approximation widely used for applications in heavy-ion collisions \cite{Dellafiore:1991zz,Blattel:1993} evaluating
the expectation value of the four-point fermion interaction in the Hartree approximation
(i.e. at mean field level). Finally only the scalar and vector components of the Wigner function are retained thanks to the spin saturated nature of the systems we are interested in.
One finally obtains the Boltzmann-Vlasov transport equations for the (anti-) quark phase-space distribution function $f^\pm$:

\begin{eqnarray}
\label{VlasovNJL}
p^{\mu}\partial_{\mu} f^{\pm}(x,p)+M(x)\partial_{\mu} M(x) \partial_{p}^{\mu} f^{\pm}(x,p)=\mathcal{C}(x,p)
\end{eqnarray}

where $\mathcal{C}(x,p)$ is the Boltzmann-like collision integral, main ingredient of the several cascade
codes already developed \cite{Molnar:2001ux,Xu:2004mz,Ferini:2008he}.
We notice that respect to the already implemented cascades the NJL dynamics
introduces a new term associated to the mass generation. Also Eq.(\ref{VlasovNJL})
is formally the same as the widely used relativistic transport approaches for hadronic matter like  RBUU,uRQMD, RLV \cite{Blattel:1993,Bass:1998ca,Fuchs:1995fa}, but with a vanishing vector field.
However a key difference is that particles do not have a fixed mass and a self-consistent derivation couples Eq.(\ref{VlasovNJL}) to the mass gap equation of the NJL model that extended
to the case of non-equilibrium can be written as:

\begin{eqnarray}
\label{NoneqMass2}
\frac{M(x)-m}{4\, g\, N_c}=M(x)\int \frac{d^3p}{(2\pi)^3} \frac{1- f^-(x,p)-f^+(x,p)}{E_p(x)}
\end{eqnarray}

and determines the local mass $M(x)$ at the space-time point $x$ in terms of the distribution functions $f^{\pm}(x,p)$.

Eqs.(\ref{VlasovNJL}) and (\ref{NoneqMass2}) form a closed system of equations constituting
the Boltzmann-Vlasov equation associated to the NJL Lagrangian that allows to
obtain self-consistently the local effective mass $M(x)$ affecting the time evolution of the distribution function $f^{\pm}(x,p)$.
A seminal work on the transport equation associated to the NJL dynamics was done in Refs.\cite{Abada:1994mf,Mishustin:1997ff},
but without a collision term, not at finite $\eta/s$ and never applied to the physical conditions of ultra-relativistic heavy-ion collisions.

For the numerical solutions of Eqs.(\ref{VlasovNJL}) and (\ref{NoneqMass2}) we use a three dimensional lattice that discretize the space
as described in Ref.\cite{Ferini:2008he,Xu:2004mz}. The standard test particle methods that sample the distribution function $f$ by mean of an ensemble of points in the phase-space is employed. The normalization condition is given by:  $ {\int d\Gamma \,f^\pm= \omega \, A (\widetilde{A}) = N_q (N_{\overline{q}})}$ with $\Gamma$ the phase space, $A_\alpha$ ($\widetilde{A}_{\alpha}$) the number of test particles (antiparticles) which are inside the considered cell and $\omega$ the proper normalization factor that relates the test particles to the real particle number.

In such a way it is possible to get a solution
of the transport equations propagating the momenta of the test (anti-)particles by mean of the relativistic Hamilton's equation. For the numerical implementation they can be written in the discretized form
as:

\begin{eqnarray}
\label{Hamiltons}
\boldsymbol{p}_i(t^+)&=&\boldsymbol{p}_i(t^-)-2 \, \delta t \, \frac{M_{\alpha}(\boldsymbol{r}_i,t)}{E_i(t)} \vec{\nabla}_r M_{\alpha}(\boldsymbol{r}_i,t) + coll. \nonumber \\
\boldsymbol{r}_i(t^+)&=&\boldsymbol{r}_i(t^-)+2 \, \delta t \,\frac{\boldsymbol{p}_i(t)}{E_i(t)}
\end{eqnarray}
with $t^\pm=t \pm \delta t$ and $\delta t$ the numerical mesh time.
The term $coll.$ on the right hand side of Eq.(\ref{Hamiltons}) indicates the effects of the collision integral as described in Ref.\cite{Xu:2004mz,Ferini:2008he}.
By mean of a reiterating procedure on time steps one gets the solutions of the transport equation
coupling Eqs.(\ref{Hamiltons}) with the gap equation, Eq.(\ref{NoneqMass2}) that discretized on lattice and for point-like test particles becomes:

\begin{eqnarray}
\label{NumericalGap}
\frac{M_{\alpha}-m}{8g N_c} = M_{\alpha}\bigg[\int_{\Lambda}\frac{d^3p}{(2\pi)^3}\frac{1}{E} -
\frac{\omega}{\Delta V_\alpha} \bigg( \sum_{i=1}^{A_{\alpha}} \frac{1}{E_i}- \sum_{i=1}^{\widetilde{A}_{\alpha}} \frac{1}{\widetilde{E}_i} \bigg) \bigg]
\end{eqnarray}

where $\Delta V_\alpha=\tau A_T  \tanh \eta_\alpha$ is the volume of each cell of the space lattice given by $A_T=0.5 fm^2$ the area in the transverse direction and $\eta_\alpha$ the space-time rapidity of the center of the cell.
The integral is instead the vacuum contribution to the gap-equation which is a divergent quantity and it is regularized by a cutoff, $\Lambda$ and has a simple analytical expression.

The space-time dependence of the mass $M_\alpha(r, t)=m-2g\big<\bar{\psi} \psi \big> $ influences the momenta of the particles because the finite gradient of the condensate generates a force which changes the momentum of a particle proportionally to $\vec{\nabla}_r \big<\bar{\psi} \psi \big>$, see Eq.(\ref{Hamiltons}).
The last is negative because the phase transition occurs earlier in the surface of the
expanding QGP fireball. Therefore
the phase transition which take place locally results in a negative contribution to the particle momenta
that makes the system more sticky respect to a free massless  gas.

\begin{figure}
\centering		
\includegraphics[scale=0.23]{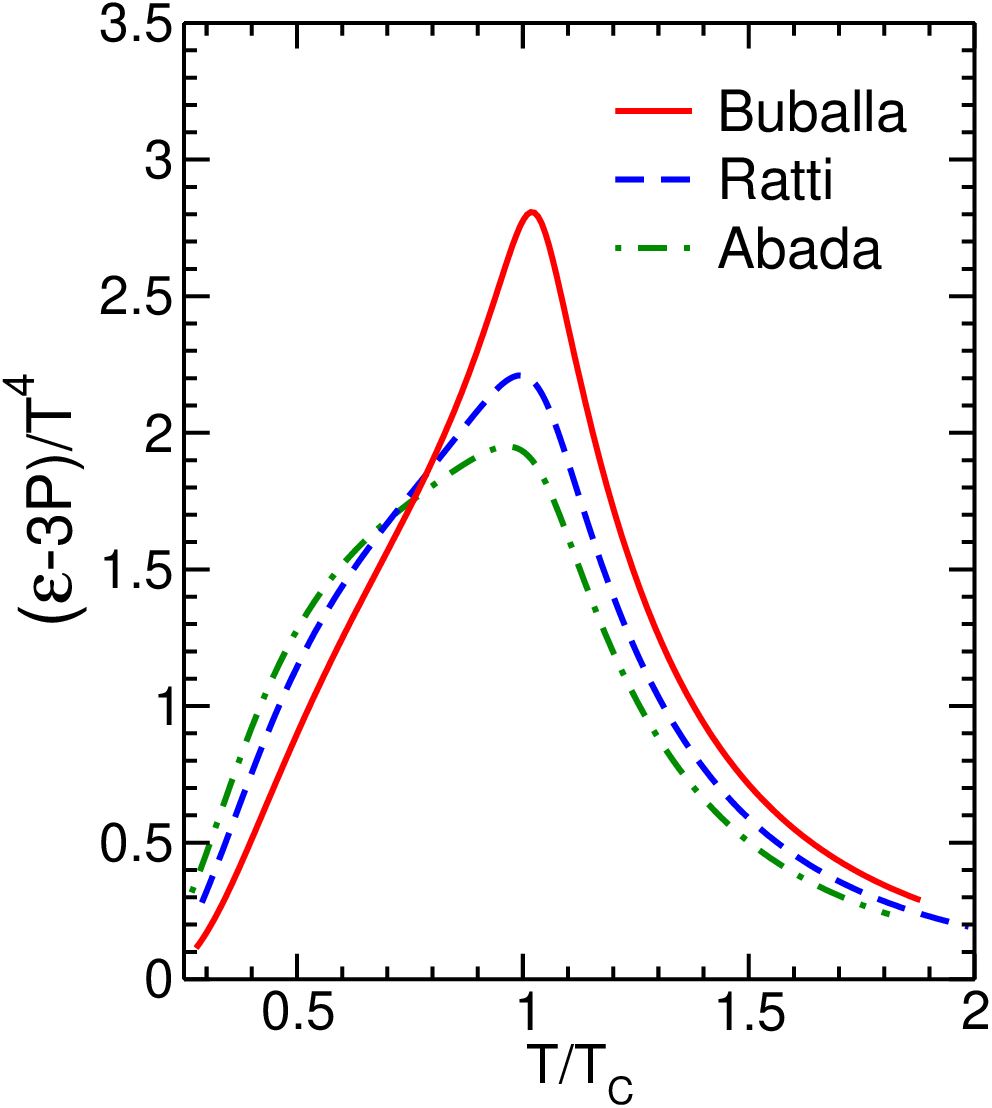}
\hspace{0.5cm}
\includegraphics[scale=0.23]{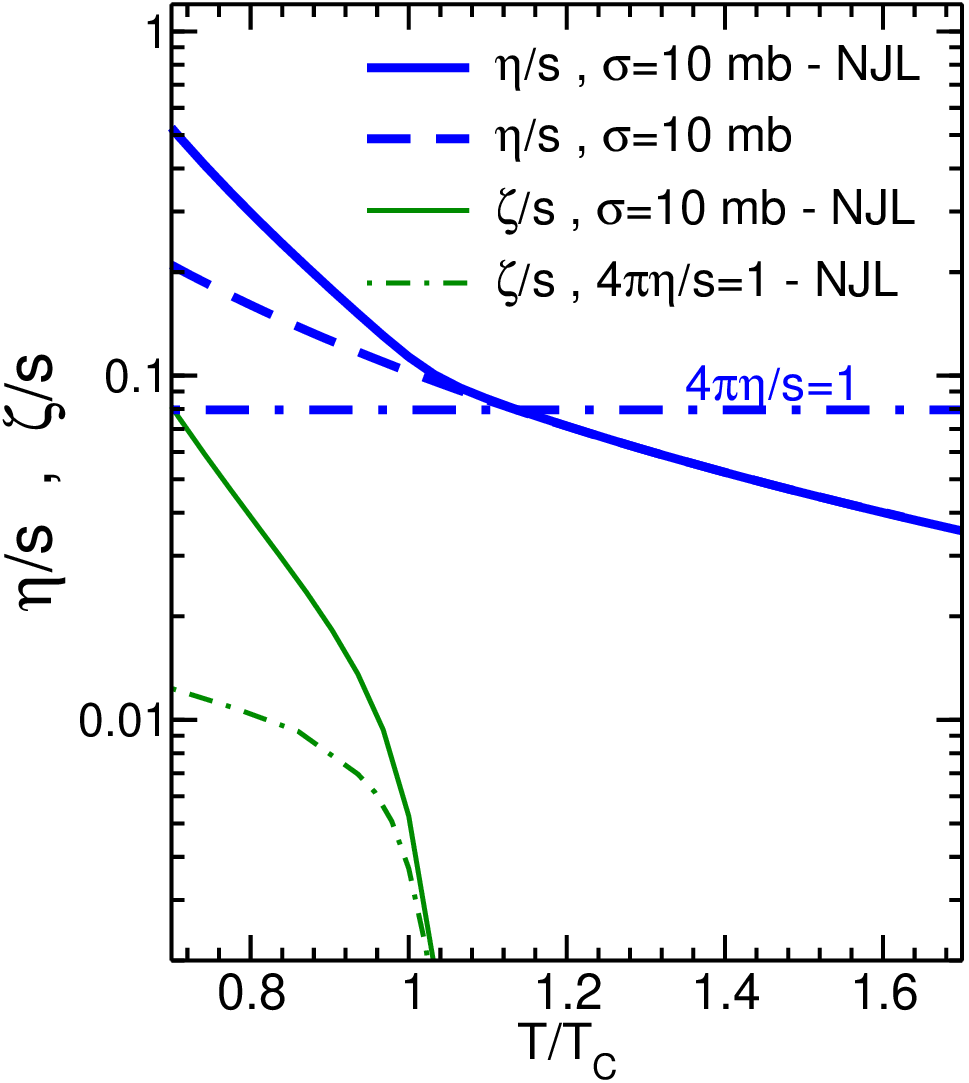}
\caption{Left: Interaction measure shown as a function of temperature T for three different
NJL parameter sets. Right: The $\eta/s$ and $\zeta/s$ for the Buballa parametrization
as a function of T for various
cases as indicated in the legend. }
\label{fig:e3p-viscosity}
\end{figure}

{\it Shear viscosity to entropy density ratio -}
The effect of the NJL mean-field can be evaluated looking at the so-called interaction measure
normalized by $T^4$, $\frac{\epsilon - 3P}{T^4}$, that gives the
deviation from the free gas relation between the energy density $\epsilon$ and the pressure $P$
that is also a measure of the breakdown of conformal symmetry. In lattice QCD it is known to be quite large
with a peak at $T_c$ and a non negligible value up to $T \sim 2-3 T_c$. In Fig.\ref{fig:e3p-viscosity} (left)
its behavior is shown for three different NJL parameterizations.
Quantitatively the QCD behavior is correctly reproduced.
In the following we will use the Buballa
parametrization that gives the largest values closer to lQCD where however the peak
is found to be about a factor two larger at $T_c$.
This means that the numerical results on the impact of the mean field in the dynamics of RHIC collisions
is reduced respect to a more realistic case.

Our final goal is to study a fluid at fixed shear viscosity to entropy ratio $\eta/s$ extending the study started at cascade level \cite{Ferini:2008he,Molnar:2008jw}.
There the strategy was to normalize locally the
cross section in order to fix the $\eta/s$ according to the simple relation $\sigma \cdot \eta/s=<p>/15n$
valid for a massless gas. Here because of the NJL field the particles acquire a mass hence both the
viscosity and the entropy density are modified respect to the simple massless case.
We briefly discuss the $\eta$ and $s$ for a system of massive particles deriving the pertinent formula to renormalize the cross section $\sigma$ in order to keep fixed locally the
$\eta/s$.
Both $\eta$ and $s$ have been derived for a thermodynamical system and has been studied also for the case of the NJL model \cite{Sasaki:2008fg}.
Here we derive expressions in terms of quantities that can be
used easily also in the numerical solution of the transport equations.
A widely used formula for $\eta$ is deduced from the relaxation time approximation, like in
Ref.\cite{Sasaki:2008fg}.
After integration by parts it is possible to write the shear viscosity for the general case of massive relativistic particles in terms of average quantities that can be easily evaluated numerically:

\begin{eqnarray}
\label{ShearNumerical}
\eta= \tau \frac{n}{15} \bigg[ 4 \:\Big< \frac{p^2}{E}\Big> + M^2\Big< \frac{p^2}{E^3}\Big>  \bigg]
\end{eqnarray}

where $\tau=\big[n \big< \sigma_{tr} v_{rel}\big>\big]^{-1}$ is the relaxation time, i.e. the time interval between two collisions, and $n$ is the total local density.
One can easily see that in the ultra-relativistic limit ($M \to 0$) the well known formula for the shear viscosity, $\eta= \frac{4}{15} \langle p\rangle/\sigma_{tr}$, is recovered.

The entropy density cannot be related simply to the local density, $s\sim 4 n$, as for the
massless case. A suitable method to evaluate locally $s$ during the dynamical evolution
of the collision is based on the use of the thermodynamical relation $sT=\epsilon + nT$
that shift the problem to the evaluation of $\epsilon,n,T$. These are easily calculated
analytically (as for Fig.\ref{fig:e3p-viscosity}) but also numerically summing up the number
of the test particles and their energy in each $\alpha-$cell.
To evaluate the temperature we exploit the general formula for a massive gas:
\begin{eqnarray}
\label{FitT}
\frac{e}{M}=\frac{3}{z} + \frac{K_1(z)}{K_2(z)}
\end{eqnarray}

with $e$ the energy per particle, $\epsilon/n$, and
$K_n(z)$ the modified Bessel functions of the second kind, $z=\frac{M}{T}$.
We know $e$ and $M$ directly from the code hence we can use Eq.(\ref{FitT}) to extract $z=\frac{M}{T}$ and therefore the temperature.
We have checked that the procedure works well performing calculation in a box with particles distributed according to a Boltzmann equilibrium distribution.

The behavior of $\eta/s$ for a thermodynamical system is shown in Fig.\ref{fig:e3p-viscosity} (right)
for a massless free gas (dashed line) and for the NJL (solid line). In both cases the cross section
is fixed to $\sigma=10\, mb$. The dot-dashed line indicates the lower bound for $\eta/s$.
We see that with a constant cross section $\sigma=10\, mb$ the $\eta/s$ is even lower than the
$1/4\pi$ at high $T$.
The lower light lines show the behavior of bulk viscosity to entropy ratio $\zeta/s$ in the NJL model
for two cases one with $\sigma=10\, mb$ (green solid line) as above and the other for
an $\eta/s$ fixed at the $1/4\pi$ value (dot-dashed line).
Of course for the massless case the bulk viscosity is zero, while for non-vanishing masses there
is a link between the $\eta$ and $\zeta$ through the relaxation time $\tau$.
The last results in a smaller growth of $\zeta/s$ when $\eta/s$ is fixed respect to the
case when $\sigma_{tr}$ is fixed.
However in both cases we can see that only
at $T<1.1\,T_c$ we have a non-vanishing $\zeta/s$. This is due to the fact that the $\zeta$
is expected to be proportional to the deviation of the sound velocity from $1/3$, $(c_s^2-1/3)^2$
that in the NJL model is known to occur only very close to $T_c$.
More importantly for our purposes is that in the NJL the $\zeta/s$ remains
order of magnitudes lower that first extrapolation from lQCD \cite{Karsch:2007jc}
and also much lower
than the smaller values used for first studies with viscous hydrodynamics \cite{Song:2009rh}.
Considering that even for much larger values of $\zeta/s$ hydrodynamics show a small effect,
we can judge safe to discard any role of $\zeta/s$ in the following results for the elliptic
flow.

We notice that
Eqs.(\ref{ShearNumerical}) and (\ref{FitT}) supply the formula for the normalization of the
cross section in each $\alpha-$cell in order to keep fixed $\eta/s$ of the system:
\begin{eqnarray}
\label{sigmarenorm}
\sigma_{tr,\alpha}= \frac{1}{15} \frac{T_\alpha}{\langle v_{rel} \rangle} \frac{4 \langle p^2/E\rangle_\alpha +
M^2_\alpha \langle p^2/E^3\rangle_\alpha}{(\epsilon_\alpha + n_\alpha T_\alpha)\, \eta/s}
\end{eqnarray}
for a massless gas $M\rightarrow 0$ ($p=E, \epsilon \sim 3nT$) and Eq.(\ref{sigmarenorm}) reduce to
the simple relation $\sigma_{tr,\alpha}=\frac{4\pi}{15}<p>_\alpha/n_\alpha$ used in Ref.s\cite{Ferini:2008he,Greco:2008fs,Molnar:2008jw,Huovinen:2008te}
for $\eta/s=1/4\pi$. Eq.(\ref{sigmarenorm}) will allow to extend such studies of a fluid at finite
viscosity to the case of partons with finite mass.

{\it Elliptic Flow - }We have run the simulations for $Au+Au$ at $\sqrt{s_{NN}}=200$ AGeV and $b=7\, fm$.
The density distribution in coordinate
space is given by the standard Glauber model. The maximum initial temperature is $T=340$ MeV and the
initial time is $\tau_0=0.6$ fm/c as usually done also in hydrodynamical calculations.
We follow the dynamical evolution of quarks, anti-quarks and gluons. The last has been included, even if they are not explicitly present in the
NJL model, with the aim of using a realistic density for both the total and the
(anti-)quark density in the simulation of the collisions.
However gluons do not actively participate in the evaluation of the
chiral phase transition, but they simply acquire the mass
of the quarks not contributing to its determination according
to the the NJL model.
The justification for this choice relies on the quasi-particle models that are fitted to lQCD thermodynamics \cite{Levai:1997yx,Castorina:2007qv}. One finds a similar behavior of $M(T)$ for both gluons and quarks approximately.
Of course for a more quantitative calculation a more careful treatment would be needed but it is not relevant to the main objective of the present seminal work, considering also that anyway the NJL model cannot be used
for an accurate quantitative study.

\begin{figure}[ht]
\vspace{0.2cm}
\includegraphics[height=1.55in,width=1.6in]{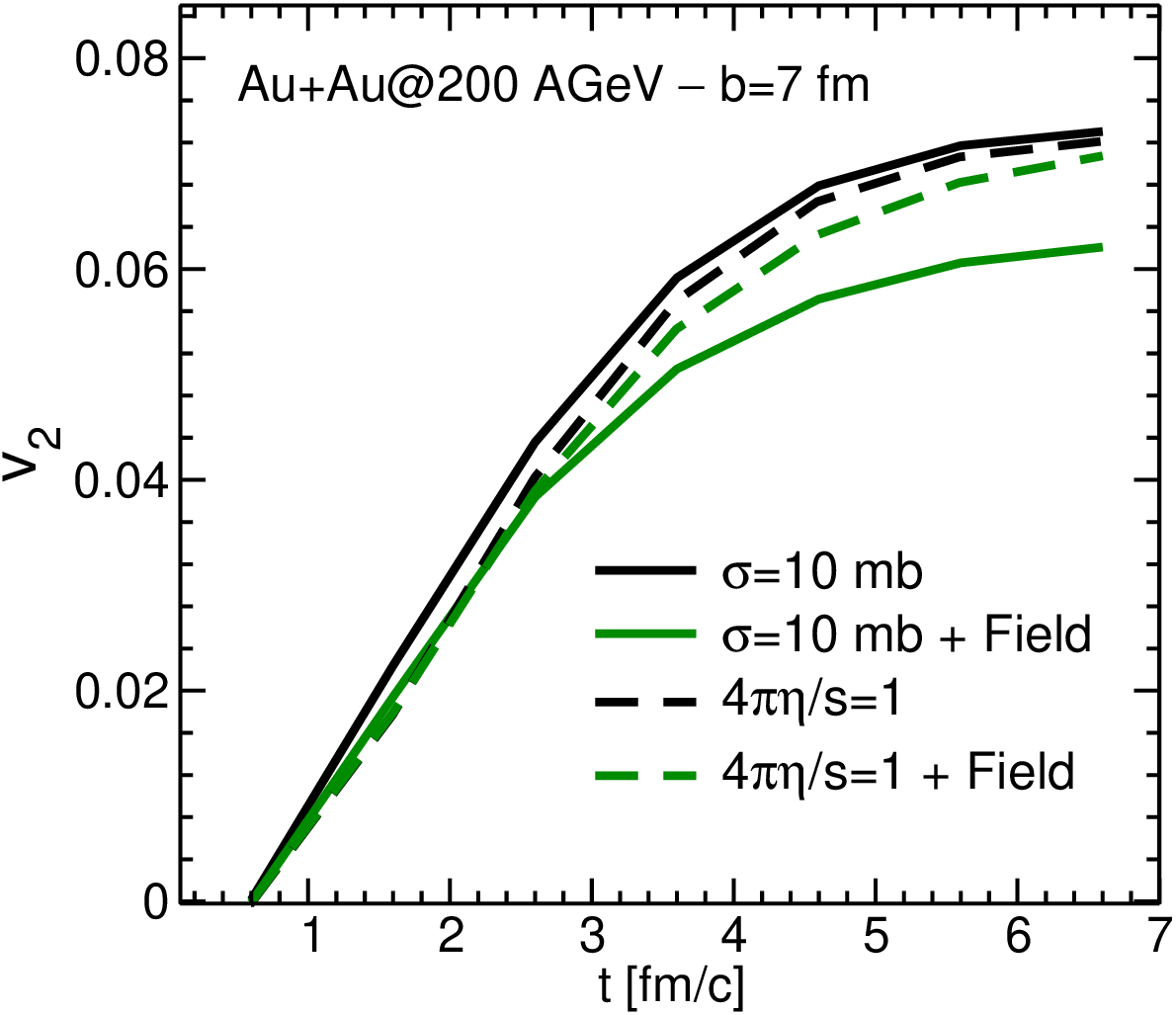}
\hspace{0.27cm}
\includegraphics[height=1.55in,width=1.6in]{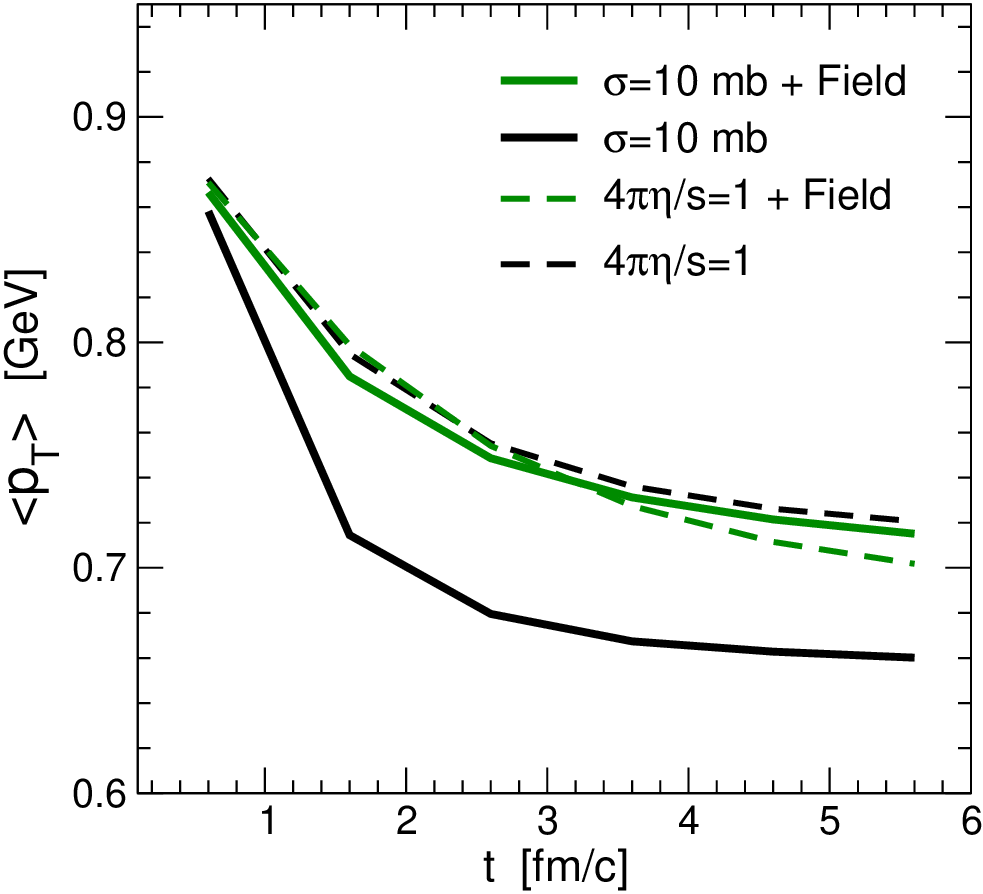}

\caption{Left: Average elliptic flow as a function of time for $Au+Au$ collisions in the mid-rapidity
region $|y|<1$ at b= 7 fm. Right: time evolution of the average transverse momentum
for the same case as in the left panel.}
\label{v2-t}
\end{figure}

\begin{figure}[ht]
\includegraphics[height=1.8in,width=2.6in]{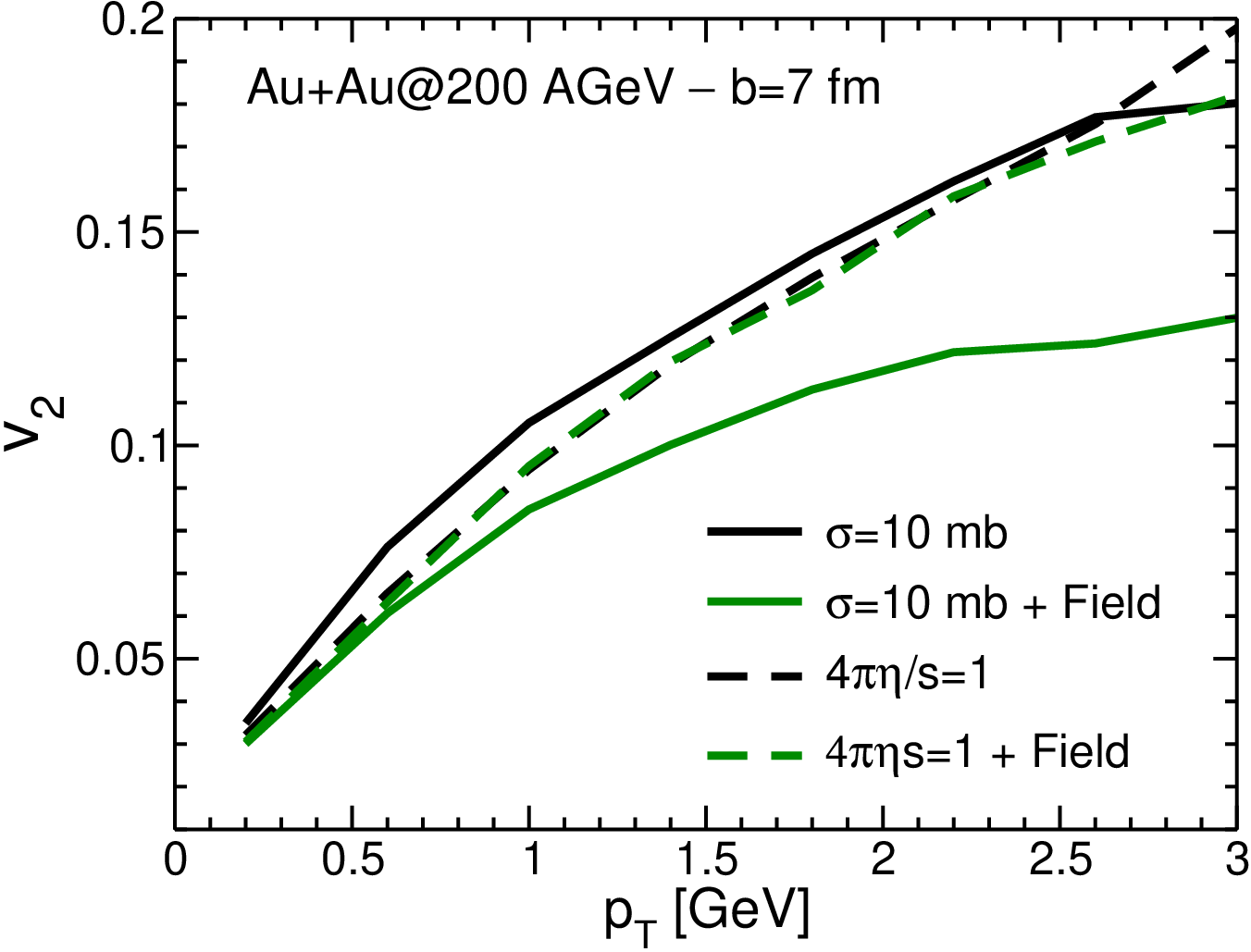}
\caption{As in Fig. 2 for the elliptic flow as a function of $p_T$.}
\label{v2-pt}
\end{figure}

In Fig. \ref{v2-t} (left) it is shown the time evolution of the average elliptic flow $\left\langle v_2\right\rangle$ for a constant transport cross section of $\sigma_{tr}=$10 mb a typical value that is able to reproduce the amount of $v_2$ observed
in experiments \cite{Lin:2001zk}. Comparing the two solid lines (black and green) we can see that the NJL mean field cause a decrease of $\left\langle v_2\right\rangle$ of about $15\%$. The reduction of $\vav$ can be expected considering that the NJL field produce
a scalar attractive field that at the phase transition results in a gas of massive particles.
In Fig.\ref{v2-pt}, we show also the elliptic flow at freeze-out as a function of the transverse momentum $p_T$.
One can see that the role of the mean field even increases with momentum affecting also particles at a
$p_T$ quite larger than the energy scale of the scalar condensate
$\langle \psi \bar\psi \rangle \sim 300 MeV$.
This is due to the fact that a
high-$p_T$ particle collides mainly with the much more abundant particles in the bulk. These
have an average momentum comparable to the strength of the scalar field: $p_T \sim 2 T \sim M_c$.
Therefore the effect of the scalar field extends thanks to collisions into
a range quite larger than one would naively think and the interplay between collisions and
mean field is fundamental.
We find that the presence of an NJL-field that drives the chiral phase transition
suppress the $v_2(p_T)$ by about $20\%$ at $p_T > 1$ GeV.
This would imply the need of a parton scattering cross section $\sigma_{tr}$ even larger than
that estimated with the cascade model which was already quite larger than the pQCD estimates
\cite{Molnar:2001ux,Lin:2004en}.
On the other hand the mean field modifies both the local entropy
density reduced by the mass generation, and the shear viscosity that increases
as shown in Fig.\ref{fig:e3p-viscosity} (right).
Therefore even if the cross section is the same with and without the NJL field the system
evolves with a different $\eta/s$.
Considering that one of the main goal is to determine the $\eta/s$ of the QGP we have investigated what
is the action of the mean field once the $\eta/s$ of the system is fixed to be the same
by mean of the cross section renormalization according to Eq. (\ref{sigmarenorm}).
Therefore we have
run simulations with and without the NJL-field but keeping constant locally the $\eta/s$.

The results for $4\pi\eta/s=1$ are shown by dashed lines in Fig.s \ref{v2-t} and \ref{v2-pt}, the (black)
dashed line is the case with only the collision term (cascade)  while the (green) dashed line is the case with the field.
We can see that once the $\eta/s$ is fixed there is essentially no difference in the calculations
with and without a field dynamics included. This is seen for both the average
$\vav$ and the $v_2(p_T)$ in Fig.s \ref{v2-t} (left) and \ref{v2-pt} respectively. This is a key result that shows that even in a microscopic approach that
distinguishes between the mean field and the collisional dynamics
the $v_2(p_T)$ is mainly driven by the $\eta/s$ of the fluid.
In other words we have found that in a microscopic approach the $\eta/s$ is the pertinent
parameter and the language of viscous hydrodynamics is appropriate. Of course this does not
mean that $v_2(p_T)$ in the transport theory is the same of the viscous hydrodynamics one, but
that, and even more importantly, the direct relation between $v_2(p_T)$ and $\eta/s$ is quite general
and their relation is not modified by the NJL field dynamics. We have checked that
this is valid also at other impact parameter ($b=3,5,9 \, fm$) and for larger
$\eta/s$ up to $\pi^{-1}$.
This is of course very important for the determination of
$\eta/s$ by mean of the data on elliptic flow and confirms the validity of the studies pursued till
now even if they miss an explicit mean field dynamics and/or the chiral phase
transition.

However we notice that while the $v_2(p_T)$ appears to be totally independent on the
presence of the NJL-field once the $\eta/s$ is kept fixed, the time evolution of $\vav$
still shows a slightly reduced elliptic flow at $t\sim 3-5 fm/c$.
A similar difference can be observed also in the time evolution of the transverse momentum
$\langle p_T \rangle$ shown in Fig. \ref{v2-t} (right).
One may ask what is the physical origin of such differences.
In principle there are two parameters affecting the $v_2(p_T)$ and the $\langle p_T \rangle$:
the sound velocity $c_s$ and the bulk viscosity $\zeta$.
As discussed previously it is safe to discard the possibility of any significant influence
of the finite $\zeta/s$ on our results considering its tiny value
in the NJL,
see Fig.\ref{fig:e3p-viscosity} (right).
It is instead reasonable that the weak decrease is due to the decrease of the
sound velocity for NJL at $T < 1.1 \,T_c$. It is well known that
$c^2_s$ decreases from $1/3$ that is the value
of a massless free gas and that this cause already in ideal hydrodynamics a
decrease of the elliptic flow \cite{Bhalerao:2005mm}. On the other hand
when the bulk of the system reaches this region most of the
$v_2(p_T)$ has already been built-up hence the effect of a moderate decrease of $c_s$ is quite weak
and could explain the small difference still visible in the time evolution of
$\vav$ and $\langle p_T \rangle$.

{\it Conclusion - }
The novelty of the present work is to be the first study within a transport approach of a fluid at finite $\eta/s$ that includes the field dynamics of the chiral phase transition.
Generally we find that at fixed cross section the effect of the NJL field is to reduce the
elliptic flow by about a $20\%$. More importantly we can state that the presence of the NJL dynamics
does not change the relation between
the elliptic flow and the $\eta/s$ that remains the same as in the cascade models
and at low $p_T$ is very close
to the one from hydrodynamics \cite{Xu:2007jv,Ferini:2008he,Molnar:2008jw,Greco:2008fs}.
If such a finding is confirmed also for a more general class of interacting quasi-particle
models it will make much safer and solid the determination of $\eta/s$
by $v_2(p_T)$. In fact as we have shown the relation is independent on the microscopic details of the interaction once the EoS and/or the $c_s^2(T)$ has been fixed.
This will be investigated in the next future, in fact the kinetic theory and the numerical implementation
presented here  can be easily extended to quasi-particle models that are fitted
to reproduce the energy density
and pressure of lQCD results. In such a case it will be possible also to study the
elliptic flow with a realistic behavior of $c_s(T)$ and  the effect of a
finite and sizable $\zeta/s$ on the elliptic flow complementing the study from viscous
hydrodynamics that are subject to problems for not too small
 $\zeta/s$ and/or for $p_T > 3T$  \cite{Song:2009rh}.

\vspace{0.35 cm}
\centerline{\bf Acknowledgements}

This work for V. Baran is supported in part by the Romanian Ministry for Education and Research
under the CNCSIS contract PNII ID-946/2007.


\end{document}